\documentclass[%
 reprint,
 amsmath,amssymb,
 aps,
]{revtex4-2}

\usepackage{graphicx}
\usepackage{dcolumn}
\usepackage{bm}
\usepackage{amsmath}
\usepackage{physics}
\usepackage{xcolor}

\begin{document}

\title{Triplet superconductivity supported by an $X_9$ high-order Van Hove singularity}

\author{Chethan Sanjeevappa $^1$}
\author{Anirudh Chandrasekaran $^{1,2}$ }
\author{Joseph J. Betouras $^1$}
\affiliation{$^1$ Department of Physics, Loughborough University, Loughborough LE11 3TU, United Kingdom \\
$^2$ Max Planck Institute for Solid State Research, D-70569 Stuttgart, Germany}


\begin{abstract}

We study a four-fold symmetric dispersion relation of a quantum material, which exhibits a single high-order Van Hove singularity of X$_9$ type at the Fermi energy. First, we analyze in detail its form, type and density of states when the energy dispersion is in its canonical form. Subsequently, we study the possibility of a superconducting state when Hubbard repulsive interactions are taken into account. By solving the gap equation, it is shown that triplet state superconductivity with power- law dependence of the critical temperature T$_c$ on the interaction strength can be formed when a single singularity is present in the Brillouin zone. We discuss the effects of fluctuations and provide an upper bound of a possible superconducting critical temperature for the ruthenate Sr$_3$Ru$_2$O$_7$ which has been shown to exhibit this type of singularity.

\end{abstract}

\maketitle

\date{\today}

\section{\label{Introduction}Introduction}

The non-trivial geometry of the electronic band structure plays a crucial role in determining the physical properties of correlated electronic phases. Foundational work by Lifshitz opened the direction of understanding the consequences of the Fermi surface geometry and the associated topological transitions, such as the formation or collapse of pockets and necks ~\cite{Lifshitz_1960, Abrikosov_1988}. These transitions can be induced by varying certain parameters in the system thereby offering greater control over the material properties. Typically, such transitions can be linked to the presence of critical points in the energy dispersion on the Fermi surface.

A critical point $\boldsymbol{k_0}$ in the energy dispersion $\varepsilon(\boldsymbol{k})$ is characterized by the vanishing Bloch velocity $\nabla \varepsilon(\boldsymbol{k_0}) = 0$ and typically corresponds to either a maximum, a minimum, or a saddle point. In two dimensions, the dispersion necessarily features a saddle point, leading to logarithmically  divergent density of states (DOS), which defines a conventional Van Hove singularity (VHS) \cite{vanHove}. In this case, the saddle is a non-degenerate point as the determinant of the Hessian is non zero $\det(\partial_{k_x} \partial_{k_y} \epsilon(\bold{k})) \neq 0$. Therefore, the energy dispersion around this point can be expanded up to the second order, leading to a canonical form $ \pm k_x^2 \mp k_y^2$. The conventional Lifshitz transitions and the associated VHS are reported in a plethora of materials including cuprates, Fe-based and ferromagnetic superconductors, heavy fermions, cobaltates and ruthenates ~\citep{Aoki, Barber, Bernhabib, Coldea, Sherkunov-Chubukov-Betouras, Slizovskiy-Chubukov-Betouras, Pfaou_etal_PRL_2013, Yelland}.

However, if the determinant of the Hessian also vanishes $\det(\partial_{k_x} \partial_{k_y} \epsilon(\bold{k})) =0$, the quadratic approximation fails, requiring an expansion of the dispersion beyond second order. Consequently, the critical points become degenerate and are referred to as higher-order critical points. As in the conventional case, the Fermi surface becomes singular and the corresponding DOS diverges, exhibiting a higher-order Van Hove singularity (HOVHS). However, HOVHSs are accompanied by a stronger power-law divergence with unique ratios for the pre-factors of energy\cite{Chandrasekaran_PRR_2020, LiangFu, Classen_Betouras_2025}. When the Fermi surface hosts HOVHSs, it can have profound consequences as they significantly amplify the electron correlations, potentially leading to interesting emergent phases. Experimentally, HOVHS can be detected as a direct signature in tunneling conductivity measurements e.g.~\cite{Yuan2019, magic_VanHove2} and the divergent DOS can lead to unusual thermodynamic and transport properties \cite{Classen_Betouras_2025}. 

With advances in experimental methods, HOVHS have attracted considerable attention in recent years~\cite{HOvHS1,HOvHS2,HOvHS3,HOvHS4,HOvHS5}. Signatures of HOVHS have been detected in several materials. Notable examples include ruthenates, such as Sr$_2$RuO$_4$, where a HOVHS has been observed at the surface \cite{Chandrasekaran_NatComms_2024}, and Sr$_3$Ru$_2$O$_7$, where a HOVHS ($X_9$) can be induced by an external magnetic field ~\cite{Efremov_PRL_2019, Chandrasekaran_PRR_2020}. Other candidate materials include overdoped graphene \cite{Rosenzweig_2020} and kagom\'e metals \cite{Kang2022, Hu2022}. In twisted bilayer graphene \cite{Cao2018, Cao2018SC}, HOVHS are linked with the appearance of almost flat bands around the magic twist angle \cite{Yuan2019}. They might also play considerable role to the understanding of the phases observed in Bernal bilayer graphene~\cite{seiler2022,Zhou_2021}. The availability of the materials hosting HOVHSs along with the ability to detect, characterize, and manipulate these singularities \cite{Chandrasekaran_NatComms_2024, Chandrasekaran_Betouras_2023}, makes this a promising research direction in controlling correlated behavior. 

Our study focuses on the X$_9$ singularity, which is directly relevant to the benchmark quantum material Sr$_3$Ru$_2$O$_7$ \cite{Grigera_Science_2001, Grigera_Science_2004, Rost_Science_2009, Rost_PNAS_2011}. $X_9$ is a fourth-order saddle of codimension 8, which is the singularity with the lowest codimension that respects four-fold symmetry. 
It does not fit into the classification scheme of the seventeen based on Thom's theorem of the Catastrophe theory \cite{Chandrasekaran_PRR_2020}, it is unimodal as we will explain below. 

We first present the $X_9$ singularity and analyze its range of influence in real systems. Then, we investigate the problem of the superconducting pairing in the presence of a single $X_9$ singularity at the Fermi surface in a system with nominally weak repulsive interactions, where we show that triplet superconductivity is possible and derive an expression for the critical temperature. We discuss the effects of fluctuations and the condition to stabilise a triplet superconducting state. We then provide an estimate of the $T_c$ for the case of Sr$_3$Ru$_2$O$_7$.
The remainder of the paper is organized as follows: in section \ref{Analysis of the singularity}, we analyze the singularity in a general framework, followed by a detailed study of superconducting pairing in section \ref{Superconductivity section}, and we conclude in section \ref{Conclusions}.

\section{\label{Analysis of the singularity}Analysis of the singularity}

\subsection{\label{general}Form and type of the singularity}

We assume a fourfold rotational symmetric dispersion expanded to quartic order, that need not have $k_x^{\,} \leftrightarrow k_y^{\,}$ reflection symmetry. This takes the form
\begin{align}
    E (\bm{k}) = & a \, k^2 + b \left( k_x^4 + k_y^4 \right) + c \left( k_x^2 \, k_y^2 \right) \nonumber\\
    & + d \left( k_x^{3} \, k_y^{\,} - k_x^{\,} \, k_y^{3} \right)
    \label{eq:dispersion}
\end{align}
Let us initially set the quadratic part to zero, i.e $a = 0$. It is easy to see that the $d$ term breaks the $k_x^{\,} \leftrightarrow k_y^{\,}$ reflection symmetry while it preserves $C_4^{\,}$ rotation taking the form $(k_x^{\,}, k_y^{\,}) \rightarrow (k_y^{\,}, -k_x^{\,})$. Equivalently, the quartic part of the dispersion takes in polar coordinates the form $k^4 \left( \beta + \gamma \, \cos 4\theta + \delta \, \sin 4\theta \right)$ with
\begin{subequations}
\begin{align}
	\beta = & \frac{6 b + c}{8}, \\
	\gamma = & \frac{2 b - c}{8}, \\
	\delta = & \frac{d}{4}.
\end{align}
\end{subequations}

Regarding the conditions for the singularity to be a higher order saddle, we expect $E (\bm{k}) = 0$ to correspond to a set of curves through the origin (rather than being a set that contains just the origin). The curves precisely contain the points where the dispersion changes sign in one direction. Since $k^4 \geqslant 0$, we need to examine the equation $\beta + \gamma \, \cos 4\theta + \delta \, \sin 4\theta = 0$. For a saddle, $\delta$ and $\gamma$ can not both be simultaneously zero since we will have a trivial maximum or minimum depending on the sign of $\beta$. With this constraint, the conditions can be rewritten as:

\begin{equation}
	 \beta + \sqrt{\gamma^2 + \delta^2} \, \cos \left( 4 \theta - \phi_0^{\,} \right) = 0, \label{eq:varphi_soln}
\end{equation}
where $\phi_0^{\,} \in [-\pi , \pi )$ is the angle for which $\cos \phi_0^{\,} = \gamma / \sqrt{\gamma^2 + \delta^2}$ and $\sin \phi_0^{\,} = \delta / \sqrt{\gamma^2 + \delta^2}$. There is a unique value of $\phi_0^{\,}$ in $[-\pi , \pi )$ and other values are of the form $2 m \pi + \phi_0^{\,}$ for integer $m$. Therefore
\begin{align}
	\phi_0^{\,} = \begin{cases}
	2 \, \tan^{-1} \left( \frac{- \gamma + \sqrt{\delta^2 + \gamma^2}}{\delta} \right), & \delta \neq 0 \\
	-\pi, & \delta = 0 \text{ and } \gamma < 0 \\
	0, & \delta = 0 \text{ and } \gamma > 0
	\end{cases}
\end{align}
Eq.~(\ref{eq:varphi_soln}) has solution only when $| \beta | \leqslant \sqrt{\gamma^2 + \delta^2}$, which is also the condition for the dispersion to become a saddle (when the inequality is strict). The solutions of Eq.~(\ref{eq:varphi_soln}) for unique values of $\theta$ modulo $2\pi$ read:
\begin{align}
	\theta_{n, \pm}^{\,} = &\frac{n \pi}{2} \pm \frac{1}{4} \cos^{-1} \left( \frac{- \beta}{\sqrt{\gamma^2 + \delta^2}} \right) + \frac{\phi_0^{\,}}{4}
\end{align}
where $n \in \{0,1,2,3\}$. These solutions for $\theta$ represent four straight lines through the origin (each $\theta_{n, \pm}^{\,}$ for $n \leqslant 1$ and its $n + 2$ partner given by $\theta_{n+2, \pm}^{\,} = \theta_{n^{\,}, \pm } + \pi$, respectively constitute the two opposite directions of a straight line emanating from the origin).

\begin{figure*}
\centering
\includegraphics[scale=1]{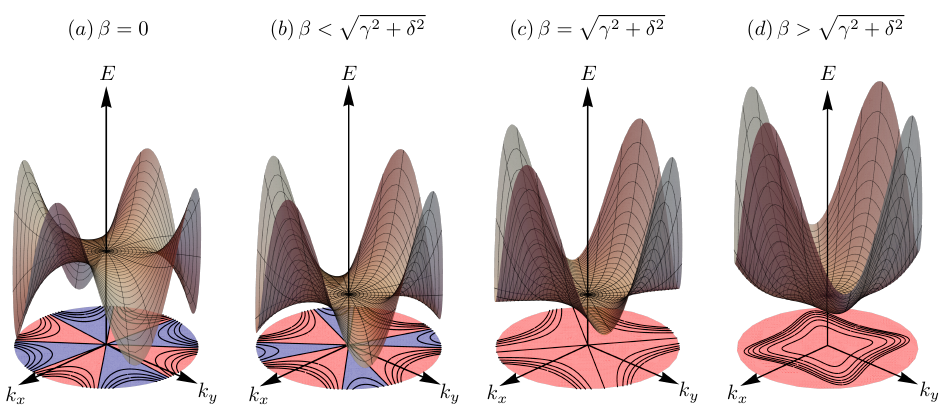}
\caption{\label{fig:cases} The general $C_4^{\,}$ symmetric $X_9^{\,}$ quartic polynomial takes the form $k^4 \left( \beta + \gamma \, \cos 4\theta + \delta \, \sin 4 \theta \right)$ in polar coordinates. As $|\beta|$ increases from a value less than $\sqrt{\delta^2 + \gamma^2}$ to greater than it, we go from a saddle to a higher order maximum/minimum (depending on the sign of $\beta$). The transition case is marked by (c) $|\beta| = \sqrt{\delta^2 + \gamma^2}$, wherein the dispersion tangentially intersects the $k$-plane along a pair of straight lines through the origin, but does not change sign. Particle and hole sectors of the $k$-plane are colored blue and red, respectively, and become equal in size in the particle-hole symmetric case, that is when (a) $\beta = 0$.}
\end{figure*}

There is a saddle when $| \beta | < \sqrt{\gamma^2 + \delta^2}$, a higher order maximum/minimum that tangentially intersects the $k_x^{\,} k_y^{\,}$ plane along two straight lines through the origin when $| \beta | = \sqrt{\gamma^2 + \delta^2}$ and a higher order maximum/minimum when $| \beta | > \sqrt{\gamma^2 + \delta^2}$. The three cases are illustrated in Fig~\ref{fig:cases}.

When the $k_x \leftrightarrow k_y$ symmetry is broken, the constant energy contours of the saddle need not be oriented symmetrically with respect to the $k_x$ and $k_y$ axes. In fact, they will be rotated by an angle with respect to the coordinate axes due to the $\phi_0$ correction appearing in the polar form of the dispersion (that is, in $\cos (4 \theta - \phi_0)$). We begin by assuming that $| \beta | < \sqrt{\gamma^2 + \delta^2}$, which guarantees a saddle. Let us first consider the case when $\delta = 0$, which ensures that the terms breaking $k_x^{\,} \leftrightarrow k_y^{\,}$ reflection are killed off, and $\phi_0^{\,} = 0$ or $\pi$. We can easily check that the zero-energy contours are straight lines that are rotated by $\pm \theta$ about the $k_x^{\,}$ and $k_y^{\,}$ axes where $\theta = \cos^{-1} \left( -\beta / \gamma \right)$. In this situation the $k_x^{\,} \leftrightarrow k_y^{\,}$ reflection is respected. When $\delta \neq 0$, we still have something analogous, with the difference that the entire system is rotated by an angle of $\phi_0^{\,}/4$. This breaks the $k_x^{\,} \leftrightarrow k_y^{\,}$ symmetry. 

\begin{figure}[h]
\includegraphics[width=\columnwidth]{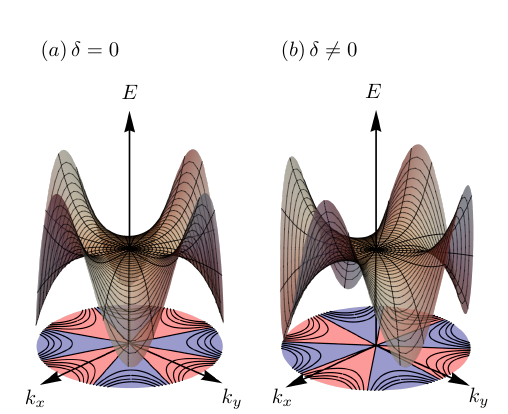}
\caption{\label{fig:delta_rotation} The $\delta$ term in $k^4 \left( \beta + \gamma \, \cos 4\theta + \delta \, \sin 4 \theta \right)$ breaks the $k_x^{\,} \leftrightarrow k_y^{\,}$ reflection symmetry. This causes a rotation of the saddle and its contours with respect to the $k_x^{\,}$ and $k_y^{\,}$ axes when $\delta \neq 0$.}
\end{figure}

\subsection{Determinacy and codimension of the $X_9^{\,}$ singularity}

First for clarity, we define the two concepts. Determinacy is the largest power of the lowest order Taylor expansion that describes the singularity. 
Codimension is the number of missing directions in polynomial space under any coordinate transformation.
For the canonical form of the $X_9^{\,}$ singularity, which is $\pm \left( k_x^4 + c \, k_x^2 k_y^2 + k_y^4 \right)$, the condition $|c| \neq 2$ ensures a finite determinacy equal to $4$ and a finite codimension equal to $8$~\cite{castrigiano}. The finite determinacy is needed to ensure that we can truncate the Taylor expansion to a finite degree (that is at least as large as the determinacy). This is allows us to then extract the DOS from the low energy dispersion~\cite{Chandrasekaran_PRR_2020}. To analyze the general, $C_4^{\,}$ symmetric form that we have been working with (defined in Eq. ~(\ref{eq:dispersion}) with $a$ set to zero), we begin with the polar form $k^4 \left( \beta + \sqrt{\gamma^2 + \delta^2} \, \cos \left( 4 \theta - \phi_0^{\,} \right) \right)$. For $\beta = 0$, by a simple rotation of the coordinate system by $\phi_0^{\,} / 4$, we can put the dispersion in the form $\sqrt{\gamma^2 + \delta^2} \left( k_x^4 - 6 \, k_x^2 k_y^2 + k_y^4 \right)$. This a particle-hole symmetric saddle with finite determinacy and codimension. 

To investigate the case $\beta \neq 0$, we first rotate the $\left( k_x^{\,}, k_y^{\,} \right)$ coordinate system with the transformation $\theta \rightarrow \theta + \left( \phi_0^{\,} + \Theta (-\beta) \, \pi \right) / 4 $, where $\Theta (x)$ is the Heaviside step function that is zero for negative $x$ and unity for positive $x$. The dispersion then reads: 
\begin{equation}
\begin{split}
E(\bm{k}) &= \operatorname{sgn}(\beta)\,k^{4}\Big(|\beta|+\sqrt{\gamma^{2}+\delta^{2}}\cos 4\theta\Big) \\
&= \operatorname{sgn}(\beta)\Big[(|\beta|+\sqrt{\gamma^{2}+\delta^{2}})(k_x^{4}+k_y^{4}) \\
&\quad +(2|\beta|-6\sqrt{\gamma^{2}+\delta^{2}})\,k_x^{2}k_y^{2}\Big].
\end{split}
\end{equation}

By rescaling $k$ by $\left( |\beta| + \sqrt{\gamma^2 + \delta^2} \right)^{1/4}$ the dispersion in the canonical form becomes:

\begin{align}
    E \left( \vb{k} \right) & = \operatorname{sgn} \left( \beta \right) \Big[ k_x^4 + k_y^4 \nonumber \\ 
    & +\left( \frac{2 |\beta| -6 \sqrt{\gamma^2 + \delta^2}}{|\beta| + \sqrt{\gamma^2 + \delta^2}} \right) k_x^2 k_y^2 \Big].
\end{align}

Thus, the condition for finite determinacy and codimension is
\begin{equation}
	\frac{ \left| 2 |\beta| -6 \sqrt{\gamma^2 + \delta^2} \right|}{|\beta| + \sqrt{\gamma^2 + \delta^2}} \neq 2
\end{equation}
This automatically rules out the previously discussed case of $|\beta| = \sqrt{\gamma^2 + \delta^2}$ which separates the maxima/minima like dispersions from saddle like ones. For a saddle with $|\beta| < \sqrt{\gamma^2 + \delta^2}$, it is easy to see that the above equation is always satisfied and this guarantees finite determinacy and codimension. Likewise, for $|\beta| > \sqrt{\gamma^2 + \delta^2}$, we will either get the trivial relation $\sqrt{\gamma^2 + \delta^2} = 0$ or the contradiction $|\beta| = \sqrt{\gamma^2 + \delta^2}$, when we set the LHS to equal $2$. The former corresponds to the condition $\delta = \gamma = 0$. Therefore, the prohibited cases are $|\beta| = \sqrt{\gamma^2 + \delta^2}$ and $\gamma = \delta = 0$. For all other values of $\beta$, $\gamma$ and $\delta$, we have a finitely determinate singularity with a finite codimension that is either a higher order maximum/minimum or a higher order saddle, depending on the magnitude of $\beta$ in comparison to $\sqrt{\delta^2 + \gamma^2}$.

It is important to note that a non-zero $\beta$ breaks particle-hole symmetry as can be seen in Fig~\ref{fig:cases}.

\subsection{Density of states in the pristine $X_9^{\,}$ saddle}
We start by analyzing the pristine, $C_4^{\,}$-symmetric $X_9^{\,}$ saddle. Note that the analysis of the maximum/minimum scenario is not necessary since it is easy to argue for a power law DOS using scaling arguments and there is no scope for a universal ratio of pre-factors since the DOS is identically zero above/below the maximum/minimum energy. Since the DOS integrals are not affected by rotating the coordinate system, we assume that we have already rotated the $\left( k_x^{\,}, k_y^{\,} \right)$ coordinate system with the transformation $\theta \rightarrow \theta + \left( \phi_0^{\,} + \Theta (-\beta) \, \pi \right) / 4 $ as before, to obtain the following dispersion relation
\begin{align}
	E \left( \vb{k} \right) =& \text{sgn} (\beta) \, k^4 \left( | \beta | + \sqrt{\gamma^2 + \delta^2} \, \cos \left( 4 \theta \right) \right) \nonumber \\
	=& \text{sgn} (\beta) \, k^4 \sqrt{\gamma^2 + \delta^2} \left( \cos \eta + \cos \left( 4 \theta \right) \right),
\end{align}
where $\eta \in (0, \pi / 2 ]$ is a parameter that characterizes the particle-hole asymmetry. When $\eta = \pi / 2$, the particle and hole sectors of the $\vb{k}$-plane are symmetric while in the $\eta \rightarrow 0$ limit, one of the sectors constricts to a set of lines rather than a two-dimensional region. To avoid cumbersome notation, let us assume $\text{sgn} (\beta) = 1$. When it is $-1$, we just have to swap the analysis of the DOS for positive and negative energies worked out below. Up to some constant prefactors, the DOS is given by

\begin{widetext}
\begin{equation}
	\nu (\epsilon) = \frac{1}{(2\pi)^2} \int\limits_0^{2\pi} d \theta \int\limits_0^{\Lambda} dk \,\, k \,\, \delta \left( \epsilon - k^4 \sqrt{\gamma^2 + \delta^2} \left( \cos \eta + \cos \left( 4 \theta \right) \right) \right)
\end{equation}

We evaluate $\nu(\epsilon)$ for both $\epsilon > 0$ and $\epsilon < 0$ cases.
Assuming that the cut-off $\Lambda$ is large enough, we set it to infinity without affecting the nature of the infrared divergence. After simplifying the integral, the expression for $\nu(\epsilon)$ in the $\epsilon > 0$ case reads:
\begin{align}
	\nu (\epsilon) =& \frac{1}{(2\pi)^2} \int\limits_0^{2\pi} d \theta \int\limits_0^{\Lambda} dk \,\, k \,\frac{ \delta \left( k - \left( \frac{\epsilon}{\sqrt{\gamma^2 + \delta^2} \left( \cos \eta + \cos \left( 4 \theta \right) \right)} \right)^{1/4} \right) }{4 \, k^3 \sqrt{\gamma^2 + \delta^2} \left( \cos \eta + \cos \left( 4 \theta \right) \right)} \, \Theta \left( \cos \eta + \cos \left( 4 \theta \right) \right) \nonumber \\
	=& \frac{|\epsilon|^{-1/2}}{16 \, \pi^2  \left( \gamma^2 + \delta^2 \right)^{1/4}} \int\limits_0^{2\pi} d \theta  \, \frac{ \Theta \left( \cos \eta + \cos \left( 4 \theta \right) \right) }{\sqrt{\cos \eta + \cos \left( 4 \theta \right)}} \text{, for } \epsilon > 0
\end{align}
\end{widetext}
The details of the calculation are in the Appendix A. If we proceed similarly for $\epsilon < 0$, then the ratio of the pre-factors of $|\epsilon|^{-1/2}$ can be expressed in a closed form in terms of the generalized hypergeometric function $_pF_q$ and the elliptic integral of the first kind $K$:
\begin{widetext}
\begin{equation}
    \frac{D_+^{\,} (\eta)}{D_-^{\,} (\eta)} = \frac{\sqrt{2 \pi } \, \eta^{-\frac{1}{2}} \left(4 \eta^2  \,
   _3F_2\left(\frac{1}{2},\frac{1}{2},1;\frac{3}{4},\frac{5}{4};\eta
   ^2\right)-\,
   _3F_2\left(\frac{3}{4},1,\frac{5}{4};\frac{3}{2},\frac{3}{2};\frac{1}{
   \eta ^2}\right) + 2 \, \eta^{\frac{3}{2}} \, \frac{1}{\sqrt{1 + \eta}} \, K\left(\frac{2}{1 + \eta }\right) \right)}{ \Gamma \left(\frac{1}{4}\right)^2 \eta \,\,
   _2F_1\left(\frac{1}{4},\frac{1}{4};\frac{1}{2};\eta ^2\right)+ 8 \,
   \Gamma \left(\frac{3}{4}\right)^2 \left(\left(1 - \eta ^2\right) \,
   _2F_1\left(\frac{3}{4},\frac{3}{4};-\frac{1}{2};\eta
   ^2\right)+\left(3 \eta ^2 - 1 \right) \,
   _2F_1\left(\frac{3}{4},\frac{3}{4};\frac{1}{2};\eta
   ^2\right)\right) }
\end{equation}
\end{widetext}

\begin{figure}[h]
\includegraphics[width=\columnwidth]{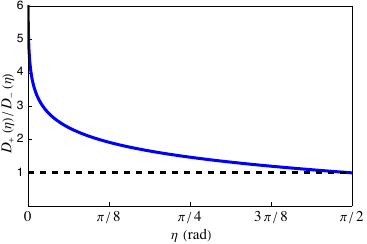}
\caption{\label{fig:DOS_ratio} The ratio of pre-factors in the power-law DOS of an $X_9^{\,}$ saddle depends on the ratio $| \beta | / \sqrt{\delta^2 + \gamma^2}$, which we define to be $\cos \eta$. When $\eta \rightarrow 0$, we expect the particle sector to shrink to zero size, which should give a diverging ratio of pre-factors. By contrast, when $\eta \rightarrow \pi / 2$, we get a particle-hole symmetric saddle that has a unit ratio of pre-factors.}
\end{figure}

The limiting value of the ratio is 1 as $\eta \rightarrow \pi / 2$ and $\infty$ as $\eta \rightarrow 0$, which is the expected behavior since $\eta = \pi / 2$ gives the particle-hole symmetric saddle and $\eta \rightarrow 0$ gives a higher order maximum that tangentially intersects the $k$-plane (as discussed earlier) which corresponds to a vanishing $D_{-}^{\,}$. The variation of the ratio with $\eta$ is shown in Fig.~(\ref{fig:DOS_ratio}).

We also find that by adding a small quadratic term, thus being off the critical tuning, we are still able to recover power law scaling of the DOS in a suitable energy range. This analysis is in the Appendix B.

\section{\label{Superconductivity section}Superconductivity}

We now study the superconducting pairing instability at a single $X_9$ Van Hove singularity at the Fermi surface. We consider a Hubbard model of interacting electrons with a nominally weak repulsive interaction U, 
\begin{align}
        \mathcal{H} = \sum_{\bm{k}\alpha} \varepsilon_{\bm{k}} c^\dagger_{\bm{k}\alpha} c_{\bm{k} \alpha} + \frac{U}{N} \sum_{\bm{k} \bm{k'} \bm{q}} c_{\bm{k + q},\uparrow} ^\dagger c_{\bm{k'-q},\downarrow}^\dagger c_{\bm{k'},\downarrow} c_{\bm{k},\uparrow}
\end{align}
where $\varepsilon_{\bm{k}}$ is the energy dispersion around the $X_9$ point and $c^\dagger_{\bm{k},\alpha}$ is a fermion creation operator for momentum $\bm{k}$ and spin $\alpha$. Further, it is assumed that the singularity is at the Fermi surface and all variables are measured from the singular point.
We choose to work with the fully symmetric form of the dispersion and use the plane polar coordinates $\varepsilon(\bm{k}) \propto k^4 \cos(4\theta)$ throughout our analysis. 

As our system involves a bare repulsive electron-electron interaction, we study the possibility of pairing driven by the Kohn-Luttinger (KL) type mechanism \cite{Kohn_luttinger, maiti_chubukov,chubukov, varma-chubukov, KL1, KL2}. 
In the KL mechanism, screening of the bare repulsion U can lead to the emergence of effective attractive channels. This process is enhanced by divergent DOS. We adopt the method developed in Ref \cite{chubukov} where the authors went beyond the one-loop Renormalization Group (RG) and study the possibility of a superconducting state by solving self-consistently the gap equation. The reason of the need to go beyond one-loop  is that due to the power-law divergence of the DOS, the corrections to RG are of the same order as the leading terms. It is worth mentioning that a Stoner analysis would suggest a ferromagnetic (Stoner) instability always, but the question is to explore possible solution of a triplet superconductivity. As a result, solving directly the gap equation and determining $T_c$, is a necessary step forward. 
This is a mean-field treatment which provides a mean-field $T_c$ and this is an important point for the discussion about fluctuations later. A general recipe though for the analysis of potential pairing as mediated by repulsive interactions is to consider the irreducible pairing vertex $\Gamma$. This irreducible pairing vertex is the anti-symmetrized interaction with zero total incoming and outgoing momenta, dressed by the renormalizations which in second order in U is given by: 

\begin{align}
    \Gamma_{\alpha \beta \gamma \nu}(\bm{k};\bm{p}) &= U (\delta_{\alpha \gamma}\delta_{\beta\nu} - \delta_{\alpha \nu}\delta_{\beta\gamma}) \nonumber\\
    &+ U^2 \Pi_{ph}(\bm{k}+\bm{p})\delta_{\alpha \gamma}\delta_{\beta\nu}     \label{eq:vertex1} \\
    &- U^2\Pi_{ph}(\bm{k}-\bm{p}) \delta_{\alpha \nu}\delta_{\beta\gamma}, \nonumber
\end{align}
where $\Pi_{ph}(\boldsymbol{k})$ is the static particle-hole polarization bubble. 

 Beyond the Random Phase Approximation (RPA) technique, the most relevant contributions are crossed diagrams (diagrams with interactions that cross each other), these are represented by insertions of particle-particle bubbles into the particle-hole channel. As previous work \cite{chubukov} shows, the restriction to second order in U of the vertex $\Gamma$  although seems questionable, is justified because the typical momenta {\bf k} and {\bf p} which are responsible for pairing, are comparable to the cutoff $\Lambda$. For such momenta, the effect of the crossed diagrams on the vertex, leading to its suppression, is small and therefore negligible. In addition, we consider that we work far from another (Stoner) instability.
To analyze the pairing problem, we employ the linearized gap equation with the vertex $\Gamma$ replacing the bare interaction U. Hence, the gap equation is given by 
\begin{equation}
\Delta_{\alpha \beta}(\bm{k}) = -\int d^2p \frac{\tanh(\frac{\epsilon_p}{2(1+Z_p) T})}{2\epsilon_p (1+Z_p)} \Gamma_{\alpha \beta \gamma \nu}(\bm{k};\bm{p}) \Delta_{\gamma \nu}(\bm{p})
\end{equation}
where $Z_p$ is the inverse quasi-particle weight that allows us to consider the self-energy corrections consistently through the relation as $\Sigma_p = i \omega Z_p$. 

The gap equation can be further decomposed into spin-singlet and spin-triplet channels,

\begin{equation}
    \Delta_{\alpha \beta}(\bm{k}) = \Delta_s(\bm{k}) i \sigma_{\alpha \beta}^y \hspace{2mm} + \Delta(\bm{k})\cdot (i \sigma^y \Vec{\sigma})_{\alpha \beta}
\end{equation}
It is obvious from Eq.~(\ref{eq:vertex1}) that the interaction is repulsive in the singlet channel and hence the associated pairing instability can be ruled out. Therefore, we search for a potential solution to the gap equation in the triplet channel. The equation in the triplet channel becomes:

\begin{align}
        \Delta_{i}(\bm{k}) =&- U^2\int d^2p \frac{\tanh(\frac{\epsilon_p}{2(1+Z_p) T})}{2\epsilon_p (1+Z_p)} \nonumber \\
        & \times \left[ \Pi_{ph}(\bm{k}+\bm{p}) - \Pi_{ph}(\bm{k} - \bm{p})\right] \Delta_i(\bm{p})
    \label{tsge}
\end{align} 
where the spin indices can be dropped in this case.
The aim is to investigate whether the gap equation can lead to a non-trivial solution with a finite critical temperature $T_c$, which is the central result of this section. 

Next, we compute the two key ingredients for this calculation, which are the static particle-hole bubble $\Pi_{ph}(\boldsymbol{q}, 0)$ and the inverse quasi-particle weight $Z_p$ at the $X_9$ point.

\subsection{Static and dynamic particle-hole bubble}

The particle-hole bubble at finite momentum and frequency transfer that is needed for the self-energy calculation reads:
\begin{align}
        \Pi_{ph}(\bm{q}, \omega_m) = \frac{1}{4\pi^2} \int \, d^2k\ \frac{f(\varepsilon_{k+q}) - f(\varepsilon_{k})}{i \omega_m + \varepsilon_{k+q} - \varepsilon_{k}}
        \label{eq:ph_general}
\end{align}
where $f$ is the Fermi function. In the limit of zero temperature and no frequency transfer, Eq.~(\ref{eq:ph_general}) reduces to 

\begin{align}
\nonumber
    \Pi_{ph}(\bm{q},0)= & \frac{1}{4\pi^2} \int \,d^2k \ \delta \left(\varepsilon(\bm{k}) \right) \\ = &
    \frac{1}{4\pi^2} \int_q^\Lambda \,dk  \ k  \   \int_0^{2\pi} \,d\theta \   \delta \left(k^4 \cos(4\theta) \right),
\end{align}
where, the cutoff $q$ captures the small- and finite-q behavior of the bubble, and $\Lambda$ is the upper cutoff for the problem. Using the standard properties of the delta function, the bubble is re-expressed as:
\begin{widetext}
\begin{equation}
     \Pi_{ph}(\bm{q}, 0) = \frac{1}{4\pi^2} \int_q^\Lambda \,dk\ \frac{k}{|k^4|} \int_0^{2\pi} \,d\theta \ \sum_{n=0}^7 \frac{\delta(\theta - \theta_n)}{|-4 \sin(4\theta_n)|}, \text{ s.t } \varepsilon(k, \theta_n) = 0,
\end{equation}
\end{widetext}
where n runs from 0 to 7 and $\theta_n$ correspond to zero dispersion lines in the patch. The final expression for the bubble becomes 
\begin{equation}
     \Pi_{ph}(\bm{q}, \omega_m = 0) = \frac{1}{4\pi^2} \left(\frac{1}{q^2} - \frac{1}{\Lambda^2}\right)
    \label{ph_static}
\end{equation}
As a caveat, we note that the polarization bubble is computed without accounting for the angular dependence. However, this approximation correctly captures the leading power of q, and including the complete angular dependence would not alter the result of our analysis in the following.


To make further progress with the dynamical polarization
bubble $\Pi_{ph}(\boldsymbol{q}, \omega_m)$, it is enough to calculate it (and subsequently the self energy) to the leading, linear order in $\omega$ as justified due to the weak interaction U. 
Therefore, Eq.~(\ref{eq:ph_general}) can be re-expressed as: 
\begin{equation}
    \Pi_{ph}(\boldsymbol{q}, \omega_m) = \frac{1}{4\pi^2} \int \,d^2k\  \frac{\delta(\varepsilon_k)}{\left(1 + \frac{i\omega_m}{\varepsilon_{k+q} - \varepsilon_k}\right)},
    \label{eq:dynamic ph general}
\end{equation}
 %
and perturbatively expand in small parameter $\omega_m/(\varepsilon_{k+q} - \varepsilon_k)$.
To linear order in $\omega_m$, the expression reads: 
\begin{align}
   \Pi_{ph}(\boldsymbol{q}, \omega_m) &\approx \frac{1}{4\pi^2} \int \, d^2k\ \delta(\varepsilon_k) \left(1 - \frac{i \omega_m}{\varepsilon_{k+q} - \varepsilon_{k}}\right)\nonumber\\
   &= \Pi_{ph}(\boldsymbol{q}, 0) + \Pi_{ph}^D(\boldsymbol{q}, \omega_m),
    \label{linear omega}
\end{align}
where $ \Pi_{ph}(\boldsymbol{q}, 0)$ is the static bubble given by Eq.~(\ref{ph_static}) and 
\begin{equation}
    \Pi_{ph}^D(\boldsymbol{q},\omega_m) \equiv \frac{-i \omega_m}{4\pi^2} \int \,d^2k\ \frac{\delta(\varepsilon_k)}{\varepsilon_{k+q} - \varepsilon_{k}}
    \label{ph bubble linear omega}
\end{equation}

The details of the calculations are presented in the Appendix C. Here we present the final result that is needed:
\begin{equation}
    \Pi_{ph}(\boldsymbol{q}, \omega_m) = \frac{1}{4\pi^2q^2}\left(1 + \frac{i\omega_m}{32q^4} \sum_{n=0}^7 (-1)^n \cot(\theta_n - \alpha)\right)
\end{equation}

\subsection{Self energy}
We are now in a position to calculate the self-energy:
\begin{equation}
    \Sigma(\omega) \sim U^2 \frac{}{}\int \,d^2q\ \int\,d\Omega\ \Pi(\boldsymbol{q}, \Omega) G(\boldsymbol{k} + \boldsymbol{q}, \Omega + \omega) 
\end{equation}
The integration variable $\Omega$ runs over the entire frequency domain, but a fixed momentum cutoff $q_{\min}$ is imposed, determined by the external frequency.

As we are primarily interested in the self-energy dependence on momentum scales rather than its detailed structure, we assume that the interacting momentum 
q is above a threshold $q_{\text{min}}$ that allows us to treat the particle-hole bubble as static \cite{chubukov}. 
This minimum momentum $q_{\text{min}}$ 
can be estimated by the inequality 
\begin{equation}
    |\omega| < \frac{ 24 q^4}{|\sum_{n=0}^7 (-1)^n\cot(\theta_n -\alpha)|}
\end{equation}
We note that $\cot(\theta_n-\alpha) \neq 0$ in its region of validity. However, there is a competing constraint Eq.~(\ref{limit angle}) that needs to be satisfied simultaneously. For completeness, we rewrite the conditions to be satisfied:
\begin{subequations}
\begin{align}
     0 < |\omega|<4\Lambda^3 q |\sin(\theta_n -\alpha)|\\
    0 < |\omega| < \frac{24 q^4}{|\sum_{n=0}^7(-1)^n \cot(\theta_n -\alpha)|}   
\end{align}  
\end{subequations}
The minimum momentum $q_{\text{min}}$ is obtained when we are closer to one of $\theta_n$, where  $\sin(x) \sim x $, and $\cot(x) \sim 1/x $. 
Therefore, to satisfy both constraints simultaneously, the lower limit can be set as 
\begin{equation}
    q_{\text{min}} \sim |\omega|^{\frac{1}{4}}
\end{equation}
By fixing the momentum scales, $q > q_{\min} \sim |\omega|^{1/4}$, the dynamic part of the bubble becomes negligible, and $\Pi_{ph}(q, \Omega) \approx \Pi_{\text{ph}}(q, 0)$ holds. Therefore the self-energy reads:

\begin{equation}
    \Sigma(\omega) =\frac{U^2}{4\pi^2} \int \,dq\ \frac{q}{q^2} \int \,d\alpha\ \int  \frac{d\Omega}{(i \Omega + i \omega -\varepsilon_{k+q})}
\end{equation}

The details of the calculation are in the Appendix D. The final result is:

\begin{equation}
    \Sigma(\omega) = i |\omega| \left( \frac{2U^2}{q^4} \log\left(\frac{\Lambda}{q}\right)  \right)
    \label{selfenergy_final}
\end{equation}
The quasiparticle weight now reads:
\begin{equation}
    Z_q = g\left(\frac{\Lambda}{q}\right)^4 \log\left(\frac{\Lambda}{q}\right),
\end{equation}
where $g \sim U^2 \Lambda^{-4}/A^2$ is the dimensionless coupling constant and A is related to the dispersion $\varepsilon_k = Ak^4 \cos(4\theta)$. The constant $g$ depends on the momentum upper cutoff explicitly. The value of A is taken to be 1 without loss of generality, throughout the calculations.

\subsection{Gap equation}

Now the central question of this work regarding the possibility of a superconducting state with a finite critical temperature, can be analyzed.
The gap equation, including the self energy, is written as:

\begin{widetext}
\begin{equation}
\Delta(\boldsymbol{k}) = \frac{U^2}{A^2} \int_{0}^\Lambda \,dp\ p \int_0^{2\pi} \,d\theta_p\ \frac{ \tanh\left(A\frac{p^4\cos(4\theta_p)}{2T_c}\right)}{2p^4\cos(4\theta_p)\left[1+g\left(\frac{\Lambda}{p}\right)^4 \log\left(\frac{\Lambda}{p}\right)\right]}  \Delta(\boldsymbol{p})\times \frac{4 k p \cos(\theta_p -\theta_k)}{k^4 + p^4 - 2k^2p^2 \cos(2\theta_p -2 \theta_k)},
\label{gap_Zp}
\end{equation}
\end{widetext}
where pre-factor $A$ is explicitly retained to keep track of the dimensions.

In principle, the gap can be a combination of all lattice harmonics that are consistent with the odd parity of the gap equation $\Delta(k, \theta_k + \pi) = -\Delta(k, \theta_k)$. However, it is not possible to analyze every channel analytically as the channels become linearly dependent. To gain insight, we analyze the gap equation in two limiting regimes $k \ll p_0$ and $k \gg p_0$, where $p_0 \sim \left(\frac{2T_c}{A}\right)^{\frac{1}{4}}$ is a natural momentum scale in the problem due to the presence of the tanh function.
In the limit $k \ll p_0$, we expand the integrand of Eq.~(\ref{gap_Zp}) in $k/p$ which is a small parameter. We retain only the term $\frac{1}{p^4}$, as it is the leading contribution that can give rise to an attractive interaction. 
In this regime, the gap is of the form $\Delta(\boldsymbol{k}) \sim k \cos\theta_k$ or $\Delta(\boldsymbol{k}) \sim k \sin\theta_k$. In the other limit $k \gg p_0$, we get a decaying form $\Delta(\boldsymbol{k}) \sim \frac{1}{k^3} \cos\theta_k$ or $\Delta(\boldsymbol{k}) \sim \frac{1}{k^3} \sin \theta_k$, suggesting a non-monotonic momentum dependence. 

In the limit $k\ll p_0$, the solution $\Delta(\boldsymbol{k}) \sim k \cos\theta_k$ or $\Delta(\boldsymbol{k}) \sim k \sin\theta_k$ belongs to the two-dimensional E representation of $C_{4v}$ group, which is the lattice version of p-wave symmetry. In principle, the gap can be a complex combination of both elements of E $(k_x,k_y)$, resulting in chiral p-wave order, which preserves $C_4$-rotational symmetry. Alternatively, the system can choose one of the directions, spontaneously breaking $C_4$ symmetry to $C_2$, leading to nematic order.  

For cases discussed above, the equation for $T_c$ reads:
\begin{widetext}
    \begin{equation}
    1 = \frac{2 U^2}{A^2} \int_0^\Lambda \,dp\ \int_0^{2\pi} \,d\theta_p\ \frac{\tanh(\frac{Ap^4 \cos(4\theta_p)}{2T_c})}{p^5 \cos(4\theta_p) \left[1 + g\left(\frac{\Lambda}{p}\right)^4 \log\left(\frac{\Lambda}{p}\right)\right]} \times \cos^2(\theta_p)
    \label{Tc_with_self_energy}
\end{equation}
\end{widetext}
To extract the dependence of $T_c$ on the coupling constant $g$ or $U$, we re-scale the momentum variables as
    $x = p \left(\frac{A}{2T_c}\right)^{\frac{1}{4}}$
which renders the integral dimensionless, and sets the upper cutoff to $ \Tilde{\Lambda} = \Lambda \left(\frac{A}{2T_c}\right)^{\frac{1}{4}}$.
%
Thus, the equation for $T_c$ becomes: 
\begin{align}
        T_c &= \frac{U^2}{A} \int_0^{\Tilde{\Lambda}} \,dx\ \int_0^{2\pi} \,d\theta_p\ \frac{\tanh(x^4 \cos(4\theta_p))}{\cos(4\theta_p)}\nonumber\\
        &\times \frac{\cos^2(\theta_p)}{x^5 + \frac{U^2}{AT_c} x \log\left(\frac{\Tilde{\Lambda}}{x}\right)} 
     \label{Tcnodim}
\end{align}
Before proceeding to the numerical solution, some comments are in order.
Regarding the convergence of the integral, in the IR region $\tanh(x^4) \approx x^4$ and the integrand roughly behaves as 
$x^3/\log\left(\frac{\Tilde{\Lambda}}{x}\right)$ which tends to zero in the limit $x \to 0$.
In the UV region, $\tanh(\Lambda^4) \to 1$ and the integrand behaves as $1/\Lambda^5$, which is negligible.
Additionally, the integrand changes sharply near $\theta_n = \frac{(2n+1)\pi}{8}$. To  
make the numerical integration accurate, we partition the domain into eight equal parts, evaluate the contribution in each domain (from $\theta_n$ to $\theta_{n+1}$) and then add all the contributions. This means, Eq.~(\ref{Tcnodim}) can be written as 
\begin{align}
         T_c &= \frac{U^2}{A} \sum_{n=0}^7 \int_0^{\Tilde{\Lambda}} \,dx\ \int_{\theta_n}^{\theta_{n+1}} \,d\theta_p\ \frac{\tanh(x^4 \cos(4\theta_p))}{\cos(4\theta_p)} \nonumber\\
         &\times \frac{\cos^2(\theta_p)}{x^5 + \frac{U^2}{AT_c} x \log\left(\frac{\Tilde{\Lambda}}{x}\right)} 
         \label{self_cons}
\end{align}
We expect $T_c$ to scale as 
\begin{equation}
    T_c \sim b\, U^\beta,
    \label{numTc}
\end{equation}
where $b$ is a constant with the appropriate dimensions. This form of $T_c$ is justified because when the interaction is switched off adiabatically, $T_c$ must decrease and vanish in the limit $U\to 0$.

\begin{figure}[h]
    \centering
    \includegraphics[width=1\linewidth]{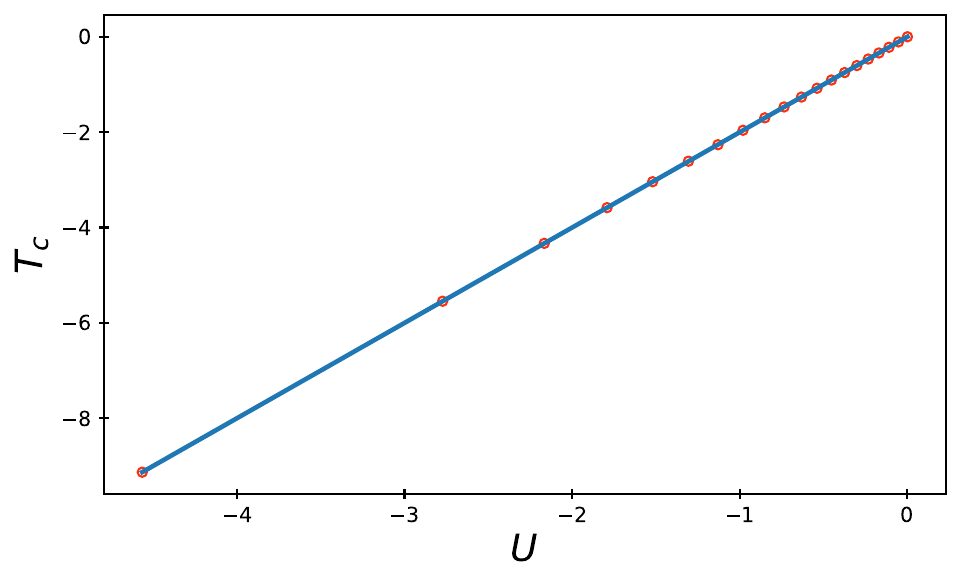}
    \caption{Log-log plot of normalized $T_c$ vs normalized U. The slope of the linear fit is 2, indicating quadratic dependence of $T_c$ on U. U ranges from $10^{-4}$ eV to $10^{-2}$ eV with A fixed at 0.1 eV.}
    \label{fig:numericTcvsU}
\end{figure}

Eq.~(\ref{self_cons}) provides a self consistent relation for $T_c$. This equation can be solved numerically using an iterative procedure for representative values of U ($0.1$ meV to $10$ meV) and A (100 meV). The resulting dependence of $T_c$ on U is found to be to a good extent quadratic, as illustrated in Fig.(\ref{fig:numericTcvsU}). This is because, the self energy correction contributes only over a very small range, whereas the dominant contribution to the integral is independent of the self energy. Consequently, the $U^2$ dependence of $T_c$ is preserved. However, the self energy naturally regularizes the integral.

\subsection{Role of fluctuations}

Regarding the role of fluctuations in connection to low dimensionality, it is clear that in purely two-dimensions we expect a Berezinskii-Kosterlitz-Thouless (BKT) transition due to proliferation of vortices with algebraic order and finite superfluid stiffness. As discussed in Ref \cite{Halperin1979}, the BKT transition temperature is then very close the transition temperature we get from the mean-field solutions of the gap equations. The reason is that the stiffness of the BKT transition is much smaller than the low temperature stiffness of a weakly coupled superconductor so that the vortex proliferation starts only very near to the mean-field $T_c$. The crucial point though is that this BKT physics holds for a triplet superconductor only in the presence of spin-orbit interaction, otherwise fluctuations destroy the mean-field solution and the BKT transition.

\section{\label{Conclusions}Discussion}

In this work we have analyzed in detail the $X_9$ singularity of a two-dimensional quantum material, which is the singularity of lowest codimension that is allowed by $\pi/2$ rotation. In the catastrophe's theory classification scheme this singularity has corank 2, codimension 8, determinacy 4 and winding number 8. Symmetry analysis suggests that the $X_9$ singularity can be realized in the band-structure of a variety of lattices at high-symmetry points. For example, at the $\Gamma$ and $M$ points of square lattices (symmetry groups $C_4$ and $D_4$). There is strong evidence that it governs the physics of the correlated material Sr$_3$Ru$_2$O$_7$ as thermodynamic measurements suggest. Therefore it presents a prime example of a HOVHS to be investigated.

We first showed that the power law scaling of the DOS can be detected in the vicinity of the critical point even if there is a small quadratic term $\propto k^2$ in the dispersion relation. This is also consistent with the study of HOVHS in the presence of disorder \cite{Chandrasekaran_Betouras_2022}. We then took into account a Hubbard repulsive interaction in the weak coupling regime, to study possible superconducting states formed in the presence of a single singularity at the Fermi surface. The central conclusion is that a triplet superconducting state can be formed, with a critical temperature that scales as a power law of the interaction strength $T_c \propto U^2$ in the weak-coupling limit.  Note that this exponent agrees with the exponent found in the numerical calculations of Ref.[{\onlinecite{chubukov}] for a different dispersion relation that provides though the same exponent, -1/2, in the divergence of the DOS. This means that the effect of the self-energy corrections is similar in both cases. It is worth emphasizing that the pairing problem has been analyzed for a single singularity. If more singularities are present within the Brillouin zone, there is always the possibility of the development of density waves \cite{HOvHS2, Zervou2023}.

Regarding Sr$_3$Ru$_2$O$_7$, where superconductivity has not been observed yet, first we note that the tuning to the $X_9$ singularity was achieved through an applied magnetic field as the tuning parameter \cite{Grigera_Science_2001, Grigera_Science_2004, Rost_Science_2009, Rost_PNAS_2011, Efremov_PRL_2019}. 
This point, in addition to the fact that there is an appreciable spin-orbit interaction in the material, makes possible the development of superconductivity and reduces the effect of the fluctuations. As a result, only a triplet superconductivity can be a solution, as we have shown in the present work.

By assuming a value of $U \approx 1.5$ meV and $A \approx 0.1$ eV, consistent with the previous studies on Sr$_3$Ru$_2$O$_7$   \cite{U_value, bandwidth, Efremov_PRL_2019} , we obtain a value of $T_c \approx 40$ mK.
In the expansion of the energy dispersion relation of the material in the vicinity of the point in the Brillouin zone where the $X_9$ is \cite{Efremov_PRL_2019}, there is an additional small quadratic term which makes the dispersion deviate from the particle-hole symmetry and the canonical form of the $X_9$ singularity. As a result, due to the fact that the estimate of the $T_c$ is made at the particle-hole symmetric point as well as the experimental conditions we have described, we can only set an upper bound on the superconducting $T_c$ through the mean-field treatment. It is evident that a possible superconducting state is detectable only at ultra-low temperatures. In the case of Sr$_3$Ru$_2$O$_7$, there is also the possibility of nesting within the Brillouin zone. Experimental results have shown at higher temperatures the formation of a SDW with a characteristic incommensurate wave-vector \cite{Lester_NatMat_2015, Efremov_PRL_2019} which connects different parts of the Fermi surface. Therefore, the estimate of the upper bound of the $T_c$ is consistent in this case.

As mentioned above, the calculation refers to the situation where a single Van Hove point is present.  More broadly, the work serves as an example for other materials that can possess a HOVHS, and especially the four-fold symmetric $X_9$, in their band structure.

\acknowledgements
We would like to thank Andrey Chubukov, Santiago Grigera, Clifford Hicks, David Perkins, Andreas Rost, Ioannis Rousochatzakis and Joerg Schmalian for useful discussions and especially Manos Kokkinis for a careful reading of the manuscript.  The work has been supported by the UK Engineering Physical Sciences Research Council grants EP/T034351/1 and EP/X012557/1.
\\

\appendix

\section{Details for the calculation of DOS for the pristine case}

Starting from Eq. 12 of the main text for $\epsilon > 0$, the integral can be represented in terms of elliptic integrals if the limits of integration are restricted carefully using the step function. In an analogous way, the integral for $\epsilon < 0$ also reads
\begin{equation}
	\nu (\epsilon) = \frac{|\epsilon|^{-1/2}}{16 \, \pi^2  \left( \gamma^2 + \delta^2 \right)^{1/4}} \int\limits_0^{2\pi} d \theta  \, \frac{ \Theta \left( - \cos \eta - \cos \left( 4 \theta \right) \right) }{\sqrt{-\cos \eta - \cos \left( 4 \theta \right)}} 
\end{equation}

The integrands in both cases are periodic with a period equal to $\pi / 2$. 
Solving the equation $\cos \eta + \cos \left( 4 \theta \right) = 0$, we obtain:
\begin{equation}
 \theta = \frac{(2 m + 1) \pi}{4} \pm \frac{\eta}{4} , m \in \mathbb{Z}   
\end{equation}
with the only distinct solutions for $\theta$ relevant being $\pi / 4 \pm \eta / 4$ as the others lie outside the integration range of $[0, \pi / 2]$ since $\eta \in ( 0, \pi/2 ]$ leading to
\begin{equation}
	0 < \frac{\pi}{4} - \frac{\eta}{4} < \frac{\pi}{4} + \frac{\eta}{4} < \frac{\pi}{2}.
\end{equation}
Thus, to compute the effect of the step functions in the integrals, we just have to examine the sign of $\cos \eta + \cos 4 \theta$ in each of the regions defined by the inequality above. For this, we just need to look at the sign of the derivative of the expression, given by $-4 \, \sin 4 \theta$. The derivative is negative at $\theta = \pi /4 - \eta / 4$ and positive at $\theta = \pi /4 + \eta / 4$. Therefore we have
\begin{widetext}
    \begin{equation}
	\text{sgn} \left( \cos \eta + \cos 4 \theta \right) = 
	\begin{cases}
		1,  & \theta \in [0, \pi/4 - \eta/4) \cup (\pi / 4 + \eta / 4, \pi / 2] \\
		-1, & \theta \in ( \pi/4 - \eta/4 , \pi / 4 + \eta / 4)
	\end{cases}
\end{equation}
Finally, we can evaluate the integrals with the restricted limits in terms of elliptic integrals $F$ and $K$ to get the ratio of pre-factors Eq. 14 of the main text.
\end{widetext}
%
%
%
%

\section{DOS with the quadratic term}
We will now include a `small' quadratic term and investigate its effect on the DOS. We will have to find some scales in the problem to quantify this smallness. A natural and obvious energy scale in the problem is $E_s^{\,} := \alpha^2 / \sqrt{\gamma^2 + \delta^2} = 8 \, a^2 / \sqrt{(2 \, b - c)^2 + 4 \, d^2}$ which in essence compares the quadratic and quartic coefficients. We will now show that for $\epsilon \ll E_s^{\,}$ the logarithmic divergence dominates while for $\epsilon \gg E_s^{\,}$, the power law regime with $-1/2$ exponent and $\eta$-dependent ratio of pre-factors dominates.
\begin{widetext}
\begin{equation}
    \nu (\epsilon) = \frac{1}{(2\pi)^2} \int\limits_0^{2\pi} d \theta \int\limits_0^{\Lambda} dk \,\, k \,\,
      \delta \left( \epsilon - \alpha \, k^2 - \text{sgn} (\beta) \, k^4 \sqrt{\gamma^2 + \delta^2} \left( \cos \eta + \cos \left( 4 \theta \right) \right) \right)
\end{equation}

We have to analyze only two distinct cases with $\text{sgn} (\alpha) \, \text{sgn} (\beta) = 1$ and $\text{sgn} (\alpha) \, \text{sgn} (\beta) = -1$. Let us assume $\text{sgn} (\beta) = 1$ so that we have to worry only about $\text{sgn} (\alpha)$ and $\text{sgn} (\epsilon)$. We can simplify the integral somewhat by the transformation $t = k^2$. Also notice the $\pi/2$ periodicity in $\theta$ of the integrand, which allows us to restrict the limits of the $\theta$ integral to $(0 , \pi / 2)$, picking up an overall factor of $4$. This then allows us to use the substitution $v = 4 \theta$, yielding a simplified version
    \begin{equation}
	\nu (\epsilon) = \frac{1}{2(2\pi)^2} \int\limits_0^{2\pi} d v \int\limits_0^{\Lambda^2} dt \,\, \delta \left( \epsilon - \alpha \, t - t^2 \sqrt{\gamma^2 + \delta^2} \left( \cos \eta + \cos v \right) \right)
\end{equation}

Let us define the following dimensionless ratio 
\begin{equation}
	r = \left( \frac{E_s^{\,}}{|\epsilon|} \right)^{1/2} = \left( \frac{\alpha^2}{|\epsilon| \, \sqrt{\gamma^2 + \delta^2}} \right)^{1/2}
\end{equation}
Our regime of interest is the `high energy' limit, that is when $r \rightarrow 0$ or equivalently $|\epsilon| \gg E_s^{\,}$. In this scenario, we can rescale the $t$ integral using the coordinate transformation $t = u \, |\epsilon|^{1/2} / (\gamma^2 + \delta^2)^{1/4}$ to obtain
    \begin{equation}
    	\nu (\epsilon) = \frac{|\epsilon|^{-1/2}}{8\pi^2 \left( \gamma^2 + \delta^2 \right)^{1/4}} \int\limits_0^{2\pi} d v \int\limits_0^{R} du \,\, \delta \left( \text{sgn}(\epsilon) - r \, \text{sgn}(\alpha) \, u - u^2 \left( \cos \eta + \cos v \right) \right),
	\label{eq:DOS_r_and_R}
\end{equation}
where we have defined another dimensionless ratio appearing in the upper limit of the $u$ integral
\begin{equation}
	R = \frac{\Lambda^2 \left( \gamma^2 + \delta^2 \right)^{1/4}}{|\epsilon|^{1/2}} = \left( \frac{\sqrt{\gamma^2 + \delta^2} \, \Lambda^4}{|\epsilon|} \right)^{1/2}.
\end{equation}
The condition that $R$ is much larger than $1$ would be tantamount to saying that the energy of interest, $\epsilon$ is well within the range where the quartic term adequately describes the dispersion. This is so because the cutoff $\Lambda$ is chosen to restrict the DOS integration domain in momentum space to the region where the quartic term is important. Therefore, $\sqrt{\gamma^2 + \delta^2} \, \Lambda^4$ precisely provides the energy equivalent of the momentum cutoff. Finally, when $E_s^{\,} \ll \epsilon \ll \sqrt{\gamma^2 + \delta^2} \, \Lambda^4$, we can ignore the small terms and obtain to a first approximation
\begin{equation}
    	\nu (\epsilon)  \approx \frac{|\epsilon|^{-1/2}}{8\pi^2 \left( \gamma^2 + \delta^2 \right)^{1/4}} \int\limits_0^{2\pi} d v \int\limits_0^{\infty} du \,\, \delta \left( \text{sgn}(\epsilon)  - u^2 \left( \cos \eta + \cos v \right) \right),
\end{equation}
\end{widetext}
This has the form already treated in Appendix A when we dealt with the DOS for the pure saddle. 
As stated in the main text, even if we are off the critical tuning, we will still be able to recover power law scaling of the DOS in a suitable energy range. 

\section{Details of the dynamic particle-hole bubble calculation}

In order to obtain a form for self energy, 
we need to know the dynamical polarization
bubble $\Pi_{ph}(\boldsymbol{q}, \omega_m)$. We restrict our calculations in obtaining the bubble and self energy to linear order in $\omega$, which is justified due to the weak interaction U. Therefore the knowledge of the leading behavior of the polarization bubble is enough. The full bubble given in the Eq.~(\ref{eq:ph_general}), can be re-expressed as: 

\begin{align}
    \Pi_{ph}(\boldsymbol{q}, \omega_m) =& \frac{1}{4\pi^2} \int \, d^2k\ \delta(\varepsilon_k) \left(\frac{\varepsilon_{k+q} - \varepsilon_k}{i\omega_m + \varepsilon_{k+q} - \varepsilon_{k}}\right)
    \label{eq:dynamic ph general}
\end{align}
To obtain the bubble to linear order in frequency, we consider a small but finite frequency, i.e.  $|\omega_m| \leq |\varepsilon_{k+q} - \varepsilon_k| \neq 0$. This approximation allows to make a perturbative expansion in the small parameter $\omega_m/(\varepsilon_{k+q} - \varepsilon_k)$.
\begin{equation}
    \Pi_{ph}(\boldsymbol{q}, \omega_m) = \frac{1}{4\pi^2} \int \,d^2k\  \frac{\delta(\varepsilon_k)}{\left(1 + \frac{i\omega_m}{\varepsilon_{k+q} - \varepsilon_k}\right)}
\end{equation}
To linear order in $\omega_m$, the expression reads 
\begin{align}
   \Pi_{ph}(\boldsymbol{q}, \omega_m) &\approx \frac{1}{4\pi^2} \int \, d^2k\ \delta(\varepsilon_k) \left(1 - \frac{i \omega_m}{\varepsilon_{k+q} - \varepsilon_{k}}\right)\nonumber\\
   &= \Pi_{ph}(\boldsymbol{q}, 0) + \Pi_{ph}^D(\boldsymbol{q}, \omega_m),
    \label{linear omega}
\end{align}
where $ \Pi_{ph}(\boldsymbol{q}, 0)$ is the static bubble given by Eq.~(\ref{ph_static}) and 
\begin{equation}
    \Pi_{ph}^D(\boldsymbol{q},\omega_m) \equiv \frac{-i \omega_m}{4\pi^2} \int \,d^2k\ \frac{\delta(\varepsilon_k)}{\varepsilon_{k+q} - \varepsilon_{k}}
    \label{ph bubble linear omega}
\end{equation}

\subsubsection{Linear in q}
To compute the dynamic bubble, we examine the Eq.~(\ref{ph bubble linear omega}) and expand the denominator to linear order in $\boldsymbol{q}$ (since $\boldsymbol{q}$ is assumed to be a small momentum relative to $\boldsymbol{k}$). Let $\theta$ and $\alpha$ be the angles made by $\boldsymbol{k}$ and $\boldsymbol{q}$ respectively. Then, the relative angle between interacting momenta is $\phi = \theta -\alpha$. Resolving $\boldsymbol{q}$ into components along $\hat{k}$ and $\hat{\theta}$, we obtain the required expression. 


\begin{align}
     \varepsilon_{k+q} - \varepsilon_{k} &= 4 k^3 q (\cos(4\theta)\cos(\theta-\alpha) \nonumber\\ 
     &+ \sin(4\theta)\sin(\theta-\alpha))
\end{align}
Evaluating the dynamic bubble, we find that 
\begin{widetext}

\begin{equation}
     \Pi_{ph}^D(\boldsymbol{q},\omega_m) = \frac{-i \omega_m}{4\pi^2} \int_q^\Lambda \,dk\ k \int_0^{2\pi}\,d\theta\ \frac{\delta(k^4 \cos(4\theta))}{4k^3q(\cos(4\theta)\cos(\theta-\alpha) + \sin(4\theta) \sin(\theta - \alpha))}
\end{equation}
Using the properties of delta function,

\begin{equation}
    \Pi_{ph}^D(\boldsymbol{q},\omega_m) = \frac{- i \omega_m}{16\pi^2}  \int_q^\Lambda \,dk\ \frac{k^{-6}}{q} \underbrace{\int_0^{2\pi}\,d\theta\ \sum_{n = 0}^7 \frac{\delta( \theta - \theta_n)/|-4 \sin(4\theta_n)|}{(\cos(4\theta)\cos(\theta-\alpha) + \sin(4\theta) \sin(\theta -\alpha)}}_I
\end{equation}

\end{widetext}
Performing the $\theta$-integration, $I$ can be evaluated to 

\begin{equation}
     I = \frac{1}{4} \sum_{n = 0}^7 \frac{(-1)^n}{ \sin(\theta_n - \alpha)}
\end{equation}
where $\cos(4\theta_n)=0$ and $\sin(4\theta_n) = (-1)^n $ are used. Expanding explicitly, it is easy to see that $I = 0$ due to pairwise cancellations. The integral $I$ is divergent for $\alpha = \theta_n$, however, the condition imposed $|\omega_m|< \varepsilon_{k+q} - \varepsilon_k \neq 0$ ensures that $\alpha \neq \theta_n$. Therefore, the value of the integral $I=0$ for any general $\alpha$. This calculation implies that $\Pi^D_{ph}(\boldsymbol{q}, \omega_m) = 0$ to linear order in $\omega_m$, when the dispersion $\varepsilon_{\boldsymbol{k} + \boldsymbol{q}}$ is expanded only to first order in $\boldsymbol{q}$ and $\theta_{\boldsymbol{k} + \boldsymbol{q}}$. As a further step, we include the second order terms in the expansion of $\varepsilon_{k+q}$. 

\subsubsection{Quadratic in q}

To second order in q, the Taylor expansion of $\varepsilon_{k+q}$ about $\boldsymbol{k}$ expressed in terms of gradient and Hessian is of the form 
    
\begin{equation}
    \varepsilon_{k+q} = \varepsilon_k + \boldsymbol{q}. \nabla\varepsilon_k + \frac{1}{2} \boldsymbol{q}^T H \boldsymbol{q} + \mathcal{O}(q^3)
\end{equation}

After a straightforward evaluation,
\begin{widetext}
\begin{equation}
    \begin{split}
         \varepsilon_{k+q} - \varepsilon_k  \approx  4k^3q(\cos(4\theta)\cos(\phi) +\sin(4\theta)\sin(\phi)) 
         + 12 k^2 q^2 \sin(4\theta) \sin(\phi) \cos(\phi)
          + 6 k^2 q^2 \cos(4\theta) \cos(2\phi)
    \end{split}
\end{equation}
The terms involving $\cos(4\theta)$ will vanish upon $\theta$-integration and therefore, are omitted from subsequent calculations. Retaining only the non-zero terms we get  
    \begin{eqnarray}
    \nonumber
    \Pi_{ph}^D(\boldsymbol{q},\omega_m) &=& \frac{-i \omega_m}{4\pi^2} \int_q^\Lambda \,dk\ k \int_0^{2\pi}\,d\theta\ \frac{\delta(k^4 \cos(4\theta))}{4k^3q \sin(4\theta) \sin(\theta-\alpha)[ 1 + 3 \frac{q}{k} \cos(\theta - \alpha) ]} \\
     &=& \frac{-i \omega_m}{16\pi^2} \int_q^\Lambda \,dk\ \frac{k^{-6}}{q} \underbrace{\int_0^{2\pi}\,d\theta\ \sum_{n = 0}^7 \frac{\delta(\theta - \theta_n)/|-4 \sin(\theta_n)|}{\sin(4\theta) \sin(\theta-\alpha)[1+3\frac{q}{k} \cos(\theta-\alpha)]}}_J
     \label{Dbubble}
\end{eqnarray}
\end{widetext}
After the $\theta$-integration, the result for J is 
\begin{equation}
    J = \frac{1}{4}\sum_{n=0}^7 \frac{1}{ \sin(4\theta_n) \sin(\theta_n - \alpha)[1 + 3 \frac{q}{k} \cos(\theta_n -\alpha)]}
\end{equation}
To analyze the nature of J, we can expand the denominator in powers of $\frac{q}{k}$ which is a small parameter. To $\mathcal{O}(q/k)$
\begin{equation}
    J = \frac{1}{4}\sum_{n = 0}^7 \left[ \frac{(-1)^n}{\sin(\theta_n -\alpha)} - \frac{3q}{k}\frac{(-1)^n \cos(\theta_n -\alpha)}{\sin(\theta_n - \alpha)}\right]
    \label{eqn J}
\end{equation}
The first term  of the Eq.~(\ref{eqn J}) was shown to vanish previously. Upon further expansion of J, the terms with even powers of $\frac{q}{k}$ go to zero and that of odd powers have the following structure 
\begin{equation}
    \left(\frac{q}{k}\right)^{(2n+1)} \cot(\theta_n - \alpha) \cos^{2n}(\theta_n-\alpha)
\end{equation}
Around  $\alpha \approx \theta_n$, which is the region of our interest, J behaves as $\cot(\theta_n - \alpha)$. Therefore, it is sufficient to retain only the $\mathcal{O}(q)$ to describe J. Substituting J in Eq.~(\ref{Dbubble}), we get 
\begin{equation}
   \Pi^D_{ph}(\boldsymbol{q}, \omega_m) = \frac{i\omega_m}{16\pi^2}\frac{3}{4} \int_q^\Lambda  \,dk\ k^{-7} \sum_{n=0}^7 (-1)^n \cot(\theta_n - \alpha) 
\end{equation}

\begin{equation}
    \Pi_{ph}^D(\boldsymbol{q}, \omega_m) = \frac{i\omega_m}{128\pi^2}\left[ \frac{1}{q^6} - \frac{1}{\Lambda^6}\right] \sum_{n=0}^7 (-1)^n \cot(\theta_n - \alpha) 
\label{dynamic bubble}
\end{equation}
Eq.~(\ref{dynamic bubble}) is valid only in the regions that satisfy $|\omega| < |\varepsilon_{k+q} - \varepsilon_k|$, which has been assumed throughout the derivations. This limits the accessibility of regions just around $\theta_n$. To impose this constraint, we examine the dispersion at $\theta_n$
\begin{align}
    \varepsilon_{k+q} - \varepsilon_k &\approx  4k^3q \sin(4\theta_n) \sin(\phi) \nonumber \\
    &+ 16k^2q^2 \sin(4\theta_n) \sin(\phi) \cos(\phi),
\end{align}
which should satisfy $|\omega| < |\varepsilon_{k+q} - \varepsilon_k|$ or equivalently 
\begin{equation}
    |\omega|< 4k^3q \sin(\theta_n - \alpha)\left[1 + 4 \frac{q}{k} \cos(\theta_n -\alpha)\right],
\end{equation}
and it is assumed that $q << k$, this allows us to estimate the constraint around the UV cut-off $k \sim \Lambda$, implying 
\begin{equation}
     |\sin(\theta_n -\alpha)| > \left(\frac{|\omega|}{4\Lambda^3 q}\right),
    \label{limit angle}
\end{equation}
which should be satisfied for all $\theta_n$. This puts a restriction on $\alpha$ in determining how close it can be to $\theta_n$ (zero-dispersion lines). In other words, the angle between the interacting momenta must be greater than the cut-off imposed in Eq.~(\ref{limit angle}) to be consistent with the assumptions made to truncate the expansion of $\Pi_{ph}(\boldsymbol{q}, \omega)$ to linear $\omega$.

Ignoring the term $\Lambda^{-6}$ in comparison with $q^{-6}$ in the Eq.~(\ref{dynamic bubble}), the full particle-hole bubble will be
\begin{equation}
    \Pi_{ph}(\boldsymbol{q}, \omega_m) = \frac{1}{4\pi^2q^2}\left(1 + \frac{i\omega}{32q^4} \sum_{n=0}^7 (-1)^n cot(\theta_n - \alpha)\right)
\end{equation}

\section{Details of the self-energy calculation}
The integration over $\Omega$ can be done using contour techniques, the function has a simple pole at $\Omega = -\omega - i \varepsilon_{k+q}$. So, constructing a contour which comprises of the real axis and a semicircular arc $C_R$ in the lower half plane enclosing the pole, we can evaluate the integral as 
\begin{align}
    \oint \frac{d\Omega}{(i \Omega + i \omega -\varepsilon_{k+q})} &=\int_{-\infty}^{\infty} \frac{d\Omega}{(i \Omega + i \omega -\varepsilon_{k+q})}\nonumber\\
    &+\int_{C_R} \frac{d\Omega}{(i \Omega + i \omega -\varepsilon_{k+q})} \nonumber\\
    & =  2\pi i Res(\Omega = -\omega -i \varepsilon_{k+q})
\end{align}
The integral around $C_R$ approaches $\pi$ which is a real constant and will be ignored as we are looking for a $\omega$-dependent self energy $\Sigma(\omega) = i \omega Z_q$, where $Z_q$ is the inverse quasiparticle weight. 
 As a result, the self energy is: 
\begin{equation}
    \Sigma(\omega) =\frac{i U^2}{2\pi} \int \,dq\ \frac{1}{q} \int \,d\alpha\ 
\end{equation}
The domain for the $\alpha$-integration is not continuous, there are forbidden regions imposed by the constraint Eq.~(\ref{limit angle}). Let $d \equiv \arcsin \left(\frac{|\omega|}{4\Lambda^3 q}\right)$ and the constraint $(|\theta_n - \alpha|) > d $ has to be satisfied simultaneously by all $\theta_n$ in $(0,\pi)$. Therefore, the valid region of integration will be 
\begin{equation}
  D = \bigcup_{n=0}^{3} \left( \theta_n + d,\; \theta_{n+1} - d \right)
\end{equation} 
where $\theta_n$'s are ordered and the region from $\pi$ to $2\pi$ provides the same contribution. Thus, the $\alpha$-integration can be evaluated to 
\begin{align}
        \int \,d \alpha\ & = 2 \left[ \alpha \right]_D = 2\pi - 16 d \nonumber\\
        &= 2\pi -16 \arcsin\left(\frac{|\omega|}{4\Lambda^3 q}\right)
\end{align}
where the factor of 2 is multiplied to account for the region $(\pi, 2\pi)$.

\begin{equation}
    \Sigma(\omega) = \frac{2iU^2}{\pi} \int_{q_{\text{min}}}^\Lambda \,dq\ \frac{1}{q} \left( \pi - 8 \arcsin\left(\frac{|\omega|}{4\Lambda^3 q}\right)\right)
\end{equation}
within the range of the integration,  $|\omega|<< 4\Lambda^3 q$ that allows us to approximate the arcsin function to linear order of its argument.
\begin{equation}
     \Sigma(\omega) \approx \frac{2iU^2}{\pi} \int_{|\omega|^{\frac{1}{4}}}^\Lambda \,dq\ \left(\frac{\pi}{q} - \frac{2|\omega|}{\Lambda^3 q^2}\right),
\end{equation}
\begin{equation}
    \Sigma(\omega) = i U^2\left( \frac{1}{2} \log\left(\frac{\Lambda^4}{|\omega|}\right)  +\frac{4|\omega|}{\pi \Lambda^3|\omega|^{\frac{1}{4}}}\right)
\end{equation}
This can be rewritten in the form:
\begin{equation}
    \Sigma(\omega) = i |\omega| \left[ \frac{U^2}{2} \frac{1}{|\omega|} \log\left(\frac{\Lambda^4}{|\omega|}\right)  +\frac{4U^2}{\pi \Lambda^3|\omega|^{\frac{1}{4}}}\right]
\end{equation}
Now using $q_{\text{min}} \approx |\omega|^{1/4}$ and ignoring the second term which scales as $1/{\Lambda}^3$, we obtain
Eq.~(\ref{selfenergy_final}) of the main text.

\bibliography{references}

\end{document}